\documentclass[lettersize,journal]{IEEEtran}
\usepackage{amsmath,amsfonts}
\usepackage{algorithmic}
\usepackage{algorithm}
\usepackage{array}
\usepackage[caption=false,font=normalsize,labelfont=sf,textfont=sf]{subfig}
\usepackage{textcomp}
\usepackage{stfloats}
\usepackage{url}
\usepackage{verbatim}
\usepackage{graphicx}
\usepackage{cite}
\usepackage{enumitem}
\setlist[itemize]{leftmargin=*}
\usepackage{tabularray}
\SetTblrInner{rowsep=3pt}
\usepackage{makecell}

\newcommand{\be}{\begin{equation}}
\newcommand{\ee}{\end{equation}}

\usepackage{stackengine}

\begin{document}

\title{Double-Directional V2V Channel Measurement using ReRoMA at 60 GHz}

\author{{Hussein Hammoud, Yuning Zhang, Zihang Cheng, Seun Sangodoyin,\\ Markus Hofer, Faruk Pasic, Thomas M. Pohl,  Radek Závorka, Ales Prokes, \\Thomas Zemen, \textit{Senior Member, IEEE},  Christoph F. Mecklenbräuker, \textit{Senior Member, IEEE}, \\ Andreas F. Molisch, \textit{Fellow, IEEE}}
\thanks{H. Hammoud, Y. Zhang, Z. Cheng, and A. F. Molisch are with the Ming Hsieh Department of Electrical and Computer Engineering, University of Southern California. S. Sangodoyin is with the Georgia Tech's School of Electrical Engineering and Computer Engineering. M. Hofer and T. Zemen are with the AIT Austrian Institute of Technology GmbH.  F. Pasic, T. Pohl, and C. F. Mecklenbräuker are with the Institute of Telecommunications, Technische Universität Wien. R. Závorka and A. Prokes is with the Department of Radio Electronics at Brno University of Technology.The work of HH, YZ, ZC, and AFM was partly funded by a grant from CALTRANS in the METRANS program, and partly by the National Science Foundation grant 2106602. The work of HM and TZ was funded within the Principal Scientist grant Dependable Wireless 6G Communication Systems (DEDICATE 6G). The work of RZ and AP was supported by the Czech Science Foundation under Lead Agency Project No. 23-04304L. }}


\maketitle

\begin{abstract}
The coordination of vehicles is a crucial element of autonomous driving, as it enhances the efficiency, convenience, and safety of road traffic. In order to fully exploit the capabilities of such coordination, communication with high data rate and low latency is required. It can be reasonably argued that millimeter-wave (mm-wave) vehicle-to-vehicle (V2V) systems are capable of fulfilling the aforementioned requirements. Nevertheless, in order to develop a system that can be deployed in real-world scenarios and to gain an understanding of the various effects of mm-wave propagation, it is necessary to perform radio propagation measurements and to derive radio channel models from them across a range of scenarios and environments. To this end, we have conducted measurement campaigns at 60\,GHz in a variety of situations, including driving in a convoy, driving in opposite direction on a six-lane road, and overtaking. These measurements employ a channel sounder based on ReRoMA, a recently introduced concept that enables the real-time measurement of dynamic double-directional radio channels. The evaluations presented herein encompass key channel parameters, including the path loss (path loss coefficient of approximately 1.9), the root mean square (RMS) delay spread (within a range of 5\,ns to 110\,ns), the angular spreads (in a range of 0.05 to 0.4), the power distribution among multipath components, and the channel stationarity time (multiple seconds).
\end{abstract}

\begin{IEEEkeywords}
channel measurements, double-directional, channel modelling, mm-wave, dynamic channels, v2v
\end{IEEEkeywords}

\section{Introduction}


Intelligent vehicles, which facilitate autonomous and semi-autonomous driving, are anticipated to shape transportation over the next two decades. While much research focuses on on-board sensors and their intelligent processing, another crucial component is vehicle-to-vehicle (V2V) communication, enabling coordinated driving maneuvers.

Inter-vehicle coordination is essential for improving traffic flow and reducing accidents. For instance, during a lane change, if a car can communicate its intention to merge to the vehicles in the target lane, it can do so with less inter-vehicle distance, reducing the need for unpredictable acceleration or deceleration. Furthermore, convoys of cars and trucks driving with small gaps, made possible by reliable V2V communication, can enhance fuel efficiency and traffic flow. From a safety perspective, V2V communication can alert drivers to obstacles and dangers beyond the ``line-of-sight" of their own sensors but detectable by other vehicles' sensors \cite{mecklenbrauker2011vehicular}. For example, a vehicle can notify others of a tree blocking a lane around a bend or a stalled vehicle that could cause a chain-reaction accident.

To enhance the integration of information from other vehicles with on-board sensors, it is beneficial to transmit raw sensor data (e.g., current lidar scans) to surrounding cars \cite{perfecto2017beyond}. This transmission must be low-latency, which precludes the use of highly efficient data compression algorithms that introduce delays. Consequently, the data rate required for these transmissions increases to 100-1000\,Mbit/s per car, exceeding current bandwidth capabilities. Thus, new applications must utilize mm-wave frequency bands, where several Gigahertz of bandwidth are available. For instance, in 2016, the FCC in the US allocated 14\,GHz of spectrum between 20 and 100\,GHz for new communication applications.

Developing new V2V systems requires a thorough understanding of the V2V propagation channel. Key metrics such as path loss, shadowing, delay dispersion, and angular spread must be known to develop reliable and efficient signaling methods. However, much of this information remains unknown for mm-wave frequencies. While some measurement campaigns have been conducted, they have not gathered all relevant data. Specifically, mm-wave systems need adaptive arrays to counteract significant free-space path loss, but most current measurements do not provide the necessary directional channel properties due to limitations in existing channel sounders. Furthermore, measurements in many critical scenarios are lacking. 

The goal of this work is to fill these gaps by (i) conducting measurement campaigns in various environments using a novel channel sounder capable of capturing the relevant information and parameters of interest, and (ii) developing channel models that align with physical reality and provide all necessary data for system development. 

\subsection{State of the Art}
Channel sounding involves exciting the propagation channel with a known signal, measuring the received signal, and extracting the ``channel response" from this data. The capabilities of the channel sounder determine the extent of information obtained; for instance, 
if the transmitter (TX) and/or receiver (RX) uses omnidirectional antennas, no information about the directions of the multipath components (MPCs) can be gathered. Narrow-band measurements, which use an excitation signal with a small bandwidth (in comparison to the center frequency), cannot provide details about MPC delays but offer a better signal-to-noise ratio, enabling a greater measurement range between TX and RX.

There is extensive literature on V2V channel measurements and modeling below 6 GHz, covering
various propagation environments, including urban, suburban, highway, and rural areas, see   and references therein. 
Such campaigns are mostly conducted by having two vehicles driving in the same or opposite directions on the same street. These campaigns typically derive average path loss and delay dispersion models from the measured results; the validity of the wide-sense-stationary uncorrelated scattering assumption has also been investigated  \cite{bernado2012validity, bernado2013delay, renaudin2013experimental}. 
Directional characteristics for such channels have been measured, e.g. in \cite{renaudin2013wideband,abbas2011directional,wang2017high}. 

    \begin{figure}[!t]
        \centering
        \includegraphics[width=3in]{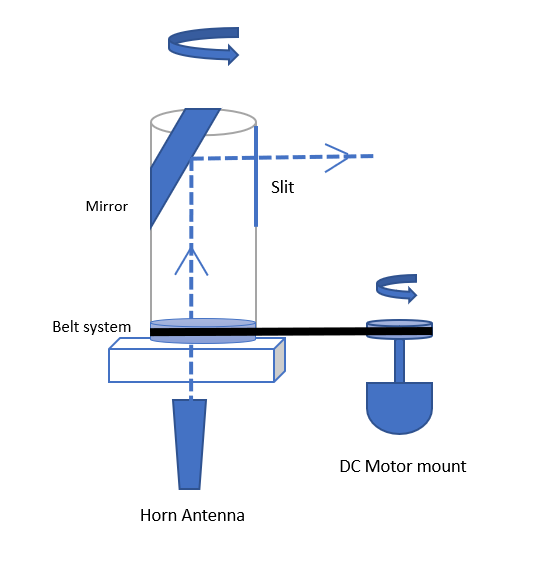}
        \caption{ReRoMA sample configuration diagram}
        \label{ContraptionDiagram}
        \end{figure}


V2V propagation channel measurements for the mm-wave band have been conducted since the early 2000s under varying conditions and using channel sounders of different capabilities. Descriptions of such measurement campaigns and modelling are available in the surveys \cite{he2019propagation,molisch2019millimeter} and references therein. Path loss, shadowing, and fading characteristics were measured primarily to model the propagation losses in the mm-wave frequency band. Many of these measurements used omnidirectional antennas \cite{ghosh2023vehicle, hoellinger2022v2v} to capture the MPCs coming from all directions. However, due to the higher isotropic free-space path loss at mm-wave frequencies, such measurements have a reduced range (maximum distance between TX and RX); furthermore such measurements do not provide any directional (angular) information.  

The measurement range can be improved by the use of directional antennas. In most measurements, horn antennas with fixed beam directions are used and may be equipped either at the TX as in  \cite{zochmann2019position, zochmann2018measured, zochmann2018statistical} (where measurements were carried out in an urban street scenario at 60\,GHz), or at the RX  \cite{hofer2022wireless} (dual-band measurements at 3.2\,GHz and 34.3\,GHz in an urban street scenario), or at both ends  \cite{wang2019comparison, wang2020shadowing} (dual-band at 2.4\,GHz and 39\,GHz in four different dynamic V2V scenarios). Other studies focused on the effect of blockage \cite{schneider2000impact, yamamoto2008path, boban2019multi, yamamoto2008path}, or modelling the small-scale fading statistics in multiple bands \cite{wang2018fading, sanchez2017millimeter}. Although these studies have been important in terms of modelling power and delay dispersion over longer distances and with better signal-to-noise ratio, they lose important information about potential MPCs reflected on surrounding structures, since MPCs arriving or departing at angles outside the horn antenna's beamwidth are not visible. 

To capture directional information, the use of antenna arrays is needed. A detailed survey of the different types of sounders is given in \cite{hammoud2024TWC}. Real arrays, which have a full RF chain for each antenna element, are cost-prohibitive at mm-wave frequencies. Switched arrays \cite{papazian2016radio} or phased arrays \cite{bas2017real, bas2019real, slezak2018effects, caudill2019phased, caudill2021phased, chopra2020real, chopra2022real} provide an attractive alternative. Still, to our knowledge, the only switched sounder used for V2V studies is the ROACH (real-time omnidirectional channel sounder) described in \cite{chopra2020real, chopra2022real, kanhere2021performance}. It has been used in multiple studies for V2V and V2X environments because of its ability to measure channels at 28 and 39\,GHz, and to capture a full MIMO snapshot in 6\,ms. However, the high cost and development effort for this sounder, made more extreme by the requirement that the sounder is rugged enough for operation while being driven on rough surfaces, is a major obstacle for building similar devices. 

The virtual array principle, where a single antenna element is mechanically translated or rotated, allows the construction of sounders with much lower cost and is thus extremely popular for mm-wave frequencies, e.g., \cite{rappaport2013millimeter,gustafson2013mm}. However, virtual arrays are only suitable for static scenarios due to the long capture duration because of the necessary mechanical movement. The main limitation for the measurement speed is the stepper motor moving the antenna. In a recent paper \cite{hammoud2024ICC,hammoud2024TWC} we presented a rotating-beam sounder that is suitable for dynamic scenarios with a capture duration of 1 s, which typically is within the stationarity time of the channel. This is achieved by employing a patent-pending redirecting rotating mirror arrangement (ReRoMA) \cite{hammoud2022method}. 
It is important to note however that such a sounder is {\em not} able to measure a MIMO snapshot within a coherence time of the channel (typically around 1 ms). Rather, ReRoMA-based sounders provide a new trade-off between cost, complexity, and measurement speed, making it an attractive method for channel sounding and modelling when switched/phased sounders are not available due to reasons of cost, power limitations, or component availability.

\subsection{Contributions}
This paper's scientific contributions are as follows:
\begin{itemize}
    \item Using our ReRoMA-based channel sounder operating at 60 GHz, we perform extensive double-directional V2V channel measurement campaigns for different driving scenarios, namely convoy, overtaking, and driving in opposite directions. 
        \item We verify evaluation results by comparing against the geometry of the environment, and discuss sample results for physical insights. 
    \item We perform statistical evaluation of the measurement data and estimate omnidirectional and directional parameters such as path loss, root-mean square (RMS) delay spread, angular spreads, MPC power distribution, and stationarity time. 
\end{itemize}

\subsection{Organization of the paper}
The remainder of the paper is organized as follows: Sec. II describes our sounder and details its characteristics. Measurement campaign scenarios are described in Sec. III, followed by the signal processing methods for the measurement evaluations in Sec. IV. Sec. V presents the results of the evaluations of the different measurement scenarios. Conclusions in Sec. VI wrap up the paper.

\section{Sounder Setup}

\subsection{Operating Principle}

    \begin{figure*}
\centering
\includegraphics[width=6in]{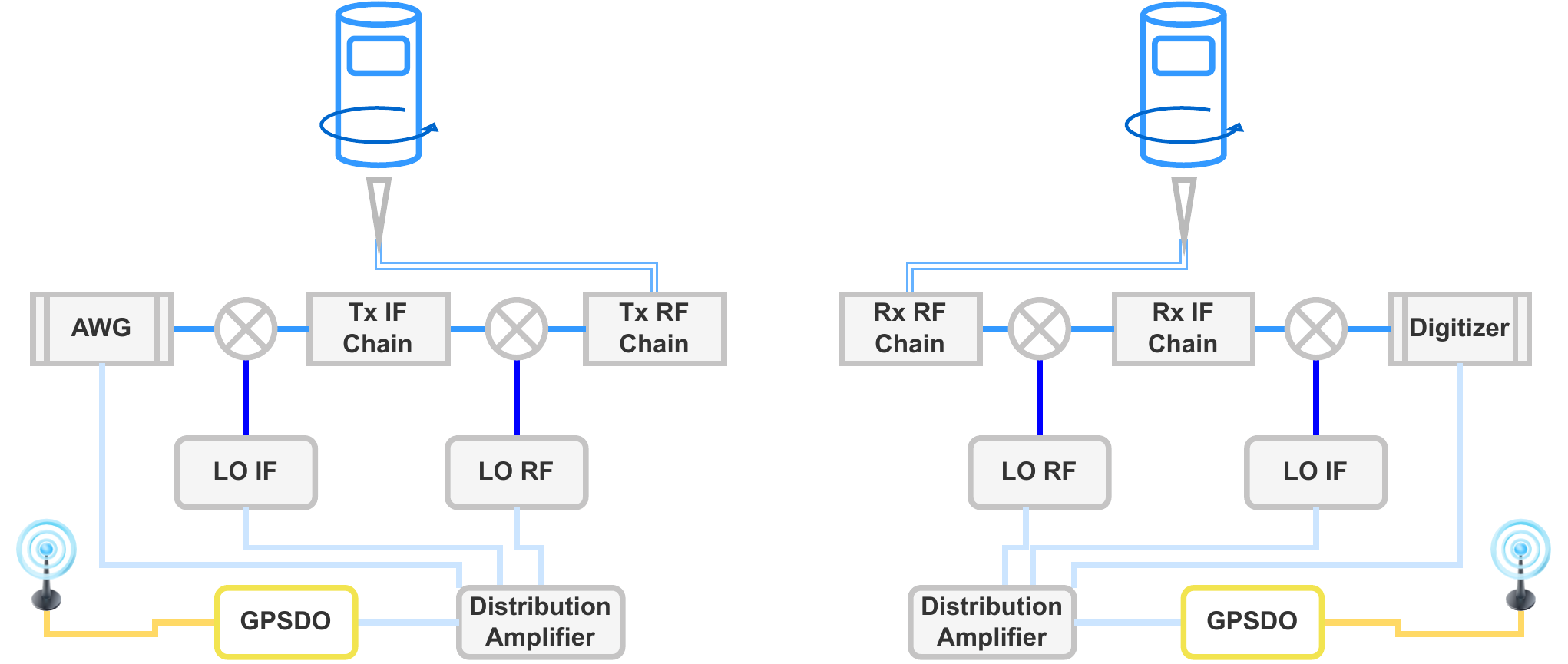}
\caption{High-level Sounder Diagram}
\label{SystemDiagram}
\end{figure*}

The mechanical structure of our sounding system is based on the patent-pending ReRoMA principle \cite{hammoud2022method, hammoud2024ICC,hammoud2024TWC} that breaks the sounder into a fixed and a rotating object. The fixed object is composed of the TX electronics, including the horn antenna, while the rotating object is based on a motor/belt/tube/mirror system that redirects the beam emitted from the horn antenna into time-varying directions, allowing the sounder to autonomously perform a comprehensive scan of the entire 360$^\circ$ azimuth plane. The re-direction is done by a periscope-style tube; the beam from the horn antenna is sent vertically into the tube, and re-directed into the horizontal plane by a mirror that is inclined at a 45 degree angle. The azimuthal direction of re-emission is determined by the orientation of the tube/mirror. The RX operates completely analoguously. A diagram for the mechanical structure in shown in Fig. \ref{ContraptionDiagram}. As none of the rotating tube's components are connected to any electronic components, rotation can be much faster than in traditional sounders where the horn (and its cable) are rotated. It is important to note that this system does not try to control directions, it only records it. Matching between directions and digitizer captures is done during the data processing phase. With such a system, we are able to capture a full ``MIMO snapshot", i.e. a combination of all 36-transmit and 72-receive directions, in about 1\,s, a duration that is typically within the stationarity time of the channel, i.e. the time-duration for which the statistics of the channel impulse response are wide-sense stationary. 

A high-level schematic of the radio frequency (RF) electronic components of the sounder can be seen in Fig. \ref{SystemDiagram}. The sounding signal pre-loaded on the arbitrary waveform generator (AWG) is a multi-tone signal that resembles a Zadoff-Chu sequences, designed to provide a flat spectrum and low peak-to-average power ratio (PAPR). This waveform is output by the AWG and upconverted to a first intermediate frequency (IF) with center frequency of 3.7\,GHz, where filtering of one of the sidebands is applied before further upconversion into the 60\,GHz band. The signal is then amplified followed by a left-hand circular-polarizer before it is radiated by a conical horn antenna. 

Finally, the signal propagates through the channel and gets captured by another conical horn antenna at the RX equipped with a polarization switch, able to alternate the circular polarization between left-hand and right-hand directions. The signal then passes through a cascade of amplifiers, bandpass filters, and mixers that downconverts it back to baseband to be captured by a National Instruments PXIe-5162 digitizer and streamed into an external harddrive for later analysis and processing. Please refer to \cite{hammoud2022method, hammoud2024ICC,hammoud2024TWC} for further details about the sounder setup.

For the measurements, the TX and RX parts of the sounder were placed on two carts, which were loaded onto the beds of two Chevrolet Silverado pickup trucks. The carts were equipped with equipment to capture location, speed, and video, which can be matched with the captured digitizer data during post-processing.


\subsection{Sounder characteristics}

A summary of all sounder parameters is given in Table \ref{tab:params}. These parameters result in a subcarrier spacing of 500\,kHz, enabling the unambiguous identification of propagation distances up to 600\,m with a delay precision of 1.5\,m.  Worth to note is that the limitations on bandwidth and angular sampling intervals were imposed by the limited backplane streaming speed of our National Instruments digitizer. 

\begin{table}[h]
    \centering
    \caption{Sounder parameters}
    \begin{tblr}{||c|c|c||}
        \hline
        Parameter & Symbol & Value \\ 
        \hline \hline 
        Number of subcarriers & ${\rm N}_{\rm sc}$ & 400 \\ \hline 
        Waveform duration & ${\rm T}_{\rm wf}$ & $2\,\mu$s \\ \hline
        RF frequency range & ${\rm f}_{\rm start}-{\rm f}_{\rm end}$ & 60.3-60.5\,GHz \\ \hline
        Measured bandwidth & ${\rm BW}$ & 200\,MHz \\ \hline
        TX Antenna 3dB Beamwitdh & $\beta_{\text{TX}}$ & $25^\circ$ \\ \hline
        RX Antenna 3dB Beamwitdh & $\beta_{\text{RX}}$ & $9^\circ$ \\ \hline
        TX/RX rotation range & [$\phi_{\rm start}:\phi_{\rm end}$)& [$-180^\circ : 180^\circ$) \\ \hline
        TX rotation resolution & $\Delta\beta_\text{TX}$ & $10^\circ$ \\ \hline
        No. of TX antenna positions & $M_\text{TX}$ & 36 \\ \hline
        RX rotation resolution & $\Delta\beta_\text{RX}$ & $5^\circ$ \\ \hline
        No. of RX antenna positions & $M_\text{RX}$ & 72 \\ \hline
        Capture trigger period & ${\rm T}_{\rm s}$ & $200\,\mu$s \\ \hline
        SIMO snapshot duration & ${\rm T}_{\rm SIMO}$ & 28\,ms \\ \hline
        MIMO snapshot duration & ${\rm T}_{\rm MIMO}$ & 1.038\,s \\ \hline
        Dynamic range & ${\rm DR}$ & 45\,dB \\ \hline
        Sampling rate & ${\rm R}_{\rm s}$ & 1.25\,GSps \\ \hline
    \end{tblr}
    \vspace{5pt}
    \label{tab:params}
    \vspace{-16pt}
\end{table}

    \begin{figure*}[!t]
        \centering
        \includegraphics[width=5in]{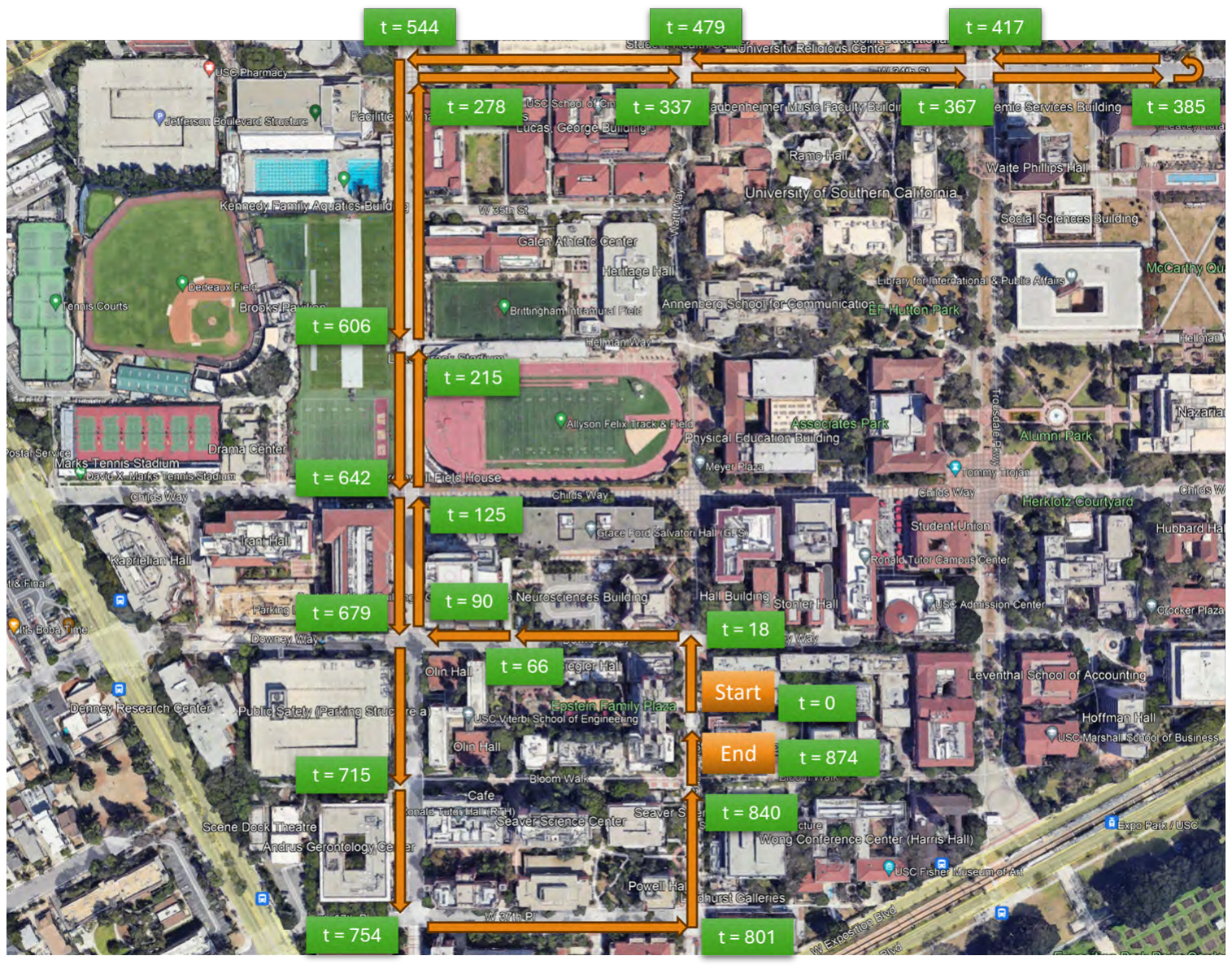}
        \caption{Convoy driving scenario: map view}
        \label{Convoy_Map}
        \end{figure*}
\section{Measurement Scenarios}
We have conducted measurement campaigns in three relevant scenarios: (i) an urban scenario where the cars, driving in a convoy, are surrounded by buildings, vegetation, and traffic, (ii) an urban scenario on a wide 6-lanes street where the cars are driving in opposite direction, (iii) an urban scenario in which the cars are overtaking each other. We describe these scenarios in more detail in the following. 

\subsection{Driving in Convoy}
This environment covered a driving track starting and ending in front of Vivian Hall of Engineering (VHE) on the USC University Park Campus (UPC), Los Angeles, CA, USA. A detailed outline of the track is shown in Fig. \ref{Convoy_Map}. The start and end points are shown in textboxes on the figure. The length of the arrows describes the distances travelled until the cars come to a full stop (i.e., at a stop sign). The environment is diverse in terms of surroundings, with some sections surrounded by vegetation, some by buildings, and some by trucks and roadwork equipment parked on the sides of the street. The operators of the trucks tried to keep a constant separation distance of 10-15\,meters while maintaining a steady driving speed of around 4\,m/s ($\simeq$ 15\,km/h). The total measurement duration was about 15\,mins, generating 874\,MIMO snapshots.

\subsection{Driving on Opposite Lanes}
\begin{figure*}[!t]
        \centering
        \includegraphics[width=5in]{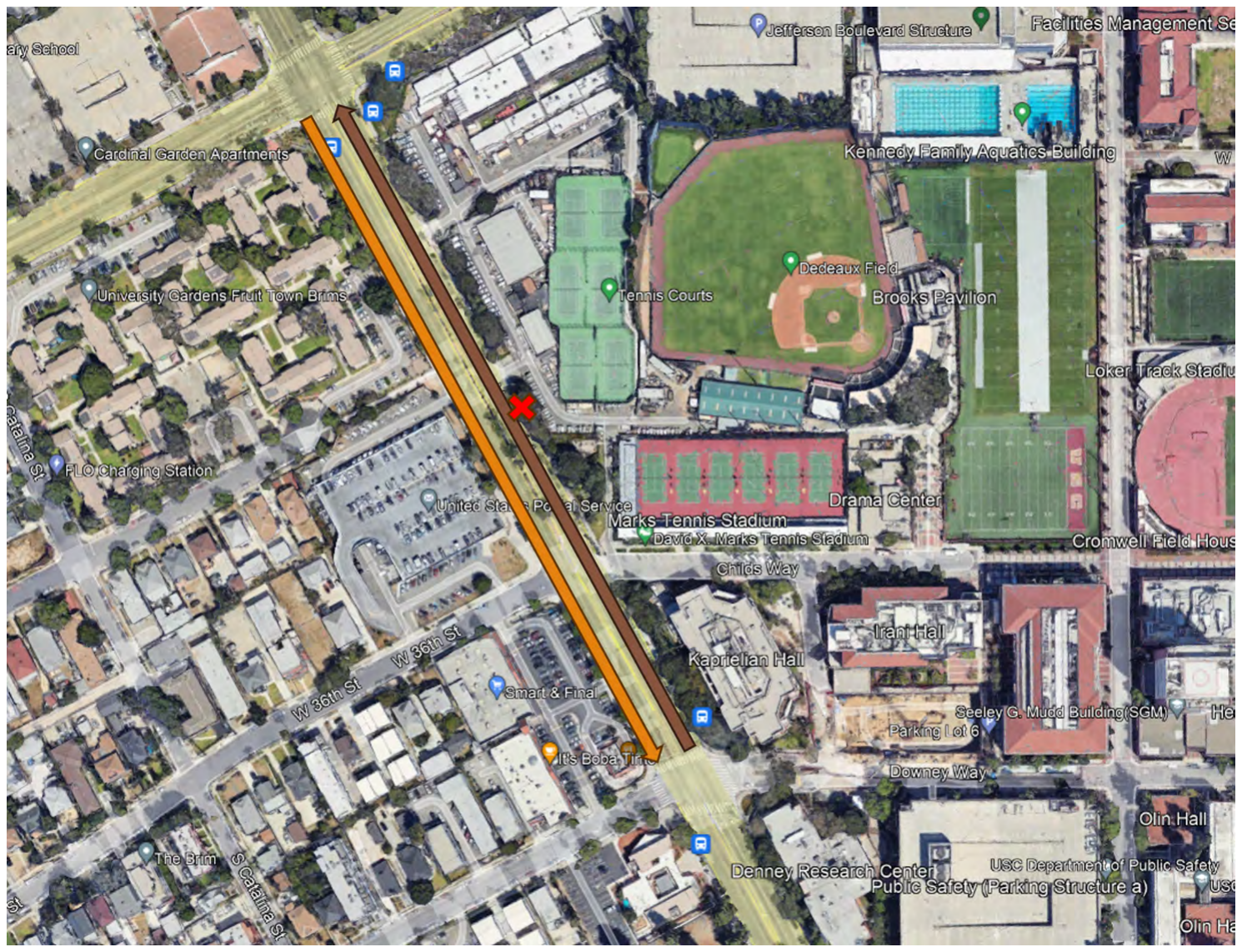}
        \caption{Opposite side and overtaking scenarios: map view. RX follows the track marked by a brown arrow for both scenarios. TX follows the track market by an orange arrow for opposite sides scenario, and is parked at the red cross location for overtaking scenario.}
        \label{other_map}
        \end{figure*}

This environment covered a section of Vermont Avenue outside of USC-UPC, stretching between the Vermont-W 36th Place and the Vermont-Jefferson intersections. A map-view of the track is shown in Fig. \ref{other_map}. The TX and RX paths are shown in orange and brown arrows respectively. The operators of the trucks synchronized the instant they started driving via walkie-talkies, driving along their respective tracks while maintaining a speed of around 4\,m/s ($\simeq$ 15\,km/h). The measurement lasted for around 2.5\,mins, generating 150\,MIMO snapshots.  

\subsection{Overtaking}
This environment covered the same section of Vermont street as the opposite-lane scenario. The TX stays stationary at the red X mark shown on the map, while the RX's path is identical to the previous scenario, and is shown as brown-colored arrow. The operators of the RX truck tried to maintain a speed of around 4\,m/s ($\simeq$ 15\,km/h) throughout the duration of the measurement. The measurement lasted for around 2.5\,mins, generating 150\,MIMO snapshots.  

\section{Evaluations Procedure}


As outlined in the preceding sections, the objective of this work is to analyze and characterize the double-directional propagation channel between the TX and the RX. Using all of the datasets captured with our equipment, we are able to organize the data of each MIMO snapshot into a three-dimensional matrix $H_{\rm meas}(f, \phi_{\rm RX}, \phi_{\rm TX})$, where $f$ represents the frequency points over the 200 MHz bandwidth, while $\phi_{\rm RX}$ and $\phi_{\rm TX}$ represent the azimuth orientations of the RX and TX beam, respectively. The dimensions of $H_{\rm meas}$ are $N_{\rm sc} \times M_{\rm TX} \times M_{\rm RX}$. The effects of the system transfer function (including the antenna within the rotating contraption) are then eliminated from the measurement data, and we obtain the calibrated directional channel transfer function $H(f, \phi_{\rm RX}, \phi_{\rm TX})$ per MIMO snapshot. For further details of the performed data processing, please refer to \cite{hammoud2024TWC}. 

The directional power delay profile (PDP) is computed as
    \begin{multline}
        P_{\rm direc}(\tau, \phi_{\rm RX}, \phi_{\rm TX})) = |\mathcal{F}^{-1}_{f} \{  H(f, \phi_{\rm RX}, \phi_{\rm TX}) \\ \cdot {\rm W}_{\rm hann}(f)\} | ^{2}
\end{multline}
where $\mathcal{F}^{-1}_{f}$ is the inverse fast Fourier transform (IFFT) with respect to $f$ and  ${\rm W}_{\rm hann}(f)$ is a Hann window applied in the frequency domain for delay-domain sidelobes suppression. We then apply noise thresholding and delay gating, similar to \cite{gomez2023impact}, as 
\begin{equation}
  P(\tau,\phi_{\rm RX}, \phi_{\rm TX}) =
    \begin{cases}
    P_{\rm direc} & \text{if $(\tau \leq \tau_{\rm gate}) \wedge (P_{\rm direc} \geq P_{\lambda})$,}\\
    0 & \text{otherwise}
    \end{cases}       
\end{equation}
where $\tau_{\rm gate}$ is the delay gating value selected to avoid incorporation of points with longer delay than what can be created in the considered environment, as well as points with ``wrap-around" effect of the IFFT, and $P_{\lambda}$ is the noise threshold. For our current measurements, $\tau_{\rm gate}$ is set to $350\,$m excess runlength, while $P_{\lambda}$ is selected to be $9\,$dB above the average noise power level of the PDP. 

To analyze the channel behavior from an ``omnidirectional" perspective, we synthesize the omni-PDPs using an approach similar to that in \cite{hur2014synchronous}. This involves reconstructing the omnidirectional pattern from the full double-directional capture by selecting the component with the maximum power (the direction with the highest contribution) for each delay bin:
 \begin{align}
 \label{eq: PLomni}
        P_{\rm omni}(\tau) = \max_{\phi_{\rm RX}, \phi_{\rm TX}} P(\tau,\phi_{\rm RX}, \phi_{\rm TX})~.
\end{align}
We also consider the PDP in the directional maximum power, 
 \begin{align}
        P_{\rm max-dir}(\tau) =  P(\tau,\phi_{\rm RX, max}, \phi_{\rm TX, max})~,
\end{align}
where $\phi_{\rm RX, max}, \phi_{\rm TX, max}$ is the direction combination providing the largest path gain as defined in (\ref{eq: PG}).

Using the omnidirectional PDPs generated for each MIMO snapshot, we proceed to compute several condensed parameters that are commonly used to characterize propagation channels, such as path gain (the reciprocal of the path loss), delay spread, angular spread, power distribution among MPCs, and stationarity time for the PDPs, see \cite[chapter 5-7]{molisch2023wireless}.

Path gain is computed as the sum of the power in each delay bin in the PDP
    \begin{align}
        PG_q = \sum_{\tau} P_{q}(\tau) ~,
        \label{eq: PG}
\end{align}
where $q$ can stand for either ``omni" or ``max-dir". 

The root-mean square (RMS) delay spread is the square root of the second central moment of the oversampled PDP
     \begin{align}
        S_{\tau,q} = \sqrt{\frac{\sum_{\tau} P_{q}(\tau)\tau^{2}}{\sum_{\tau} P_{q}(\tau)} -  T^{2}} ~~~{\rm where} ~~~T = \frac{\sum_{\tau} P_{q}(\tau)\tau}{\sum_{\tau} P_{q}(\tau)} ~.
        \label{eq: rms delay spread omni} 
    \end{align}

The angular spread gives a condensed quantification of the angular dispersion in the channel. We first generate the angular power spectrum (APS) at both the TX and the RX sides as 
\begin{align}
    {\rm APS}_{\rm TX}(\phi_{\rm TX}) = \sum_{\Tilde{\tau} \in \textbf{A}_{\phi_{\rm TX}} } P_{\rm omni} (\Tilde{\tau}) \\
    {\rm APS}_{\rm RX}(\phi_{\rm RX}) = \sum_{\Tilde{\tau} \in \textbf{B}_{\phi_{\rm RX}} } P_{\rm omni} (\Tilde{\tau}) ~.
\end{align}
where $\textbf{A}_{\phi_{\rm TX}}$ and $\textbf{B}_{\phi_{\rm RX}}$ are the set of $\tau$ values for which the TX direction $\phi_{\rm TX}$ and the RX direction $\phi_{\rm RX}$ were respectively selected to construct $P_{\rm omni}$ according to \ref{eq: PLomni}. 

It is important to note that performing noise thresholding and delay gating of $P(\tau, \phi_{\rm TX}, \phi_{\rm RX})$ is vital for suppressing the impact of noise on the delay spreads and angular spreads \cite{gomez2023impact}. 

We then proceed to compute the angular spreads from the APS by applying Fleury's definition \cite{fleury2000first} as 
    \begin{multline}
        S_{\phi} = \sqrt{\frac{\sum_{\phi} {|{\rm exp}(j\phi)-\mu_{\phi}|}^2 APS(\phi)}{\sum_{\phi} APS(\phi)}} \\ ~~~{\rm with} ~~~ \mu_{\phi} = \frac{\sum_{\phi} {\rm exp}(j\phi) APS(\phi)}{\sum_{\phi} APS(\phi)}~.
        \label{eq: angular spread without unit}
    \end{multline}
The angular spreads following this definition will be unit-less, ranging between 0 and 1, where 0 corresponds to no spread in the angular domain, and 1 corresponds to power coming uniformly from all directions. 

The power distribution over MPCs is another very important parameter for the channel analysis as it captures the ratio of the strongest MPC versus the other MPCs present in the channel. We define a parameter $\kappa$ as 
    \begin{align}
        \kappa = \frac{P_{\rm omni}(\Tilde{\tau}_{1})}{\sum_{\Tilde{\tau}=\Tilde{\tau}_{2}}^{\Tilde{\tau}_{N}} P_{\rm omni}(\Tilde{\tau})}
        \label{eq: Kappaomni}
\end{align}
where $\Tilde{\tau}_{k}$ is the location of the $k$-th local maximum (out of N total local maximas) of the omnidirectional PDP $P_{\rm omni}(\Tilde{\tau})$, ordered by magnitude, so that $\Tilde{\tau}_{1}$ signifies the location of the largest local maximum. Note that while this parameter has some similarity to the Rice factor, it is {\em not} identical, since (due to the limited resolution of the setup) we cannot extract the strongest MPC, but rather determine the magnitude of PDP peaks.  

The stationarity of the PDP is measured in units of MIMO snapshots where we compute the cross-correlation between subsequent PDPs and we mark them as part of the same stationarity region if the correlation is above a certain threshold $\alpha$. We mark the end of a region and the start of a new one whenever the correlation falls below $\alpha$, and we calculate the length of each region based on how many MIMO snapshots it contains. For our analysis, we have considered stationarity region results for $\alpha$ = 0.7 and 0.9. Under that principle, we evaluate the stationarity regions for both omnidirectional and max-dir PDPs. We additionally distinguish between the PDPs with/without time-of-arrival (TOA)-normalization to the LOS component for both omni and max-dir cases. TOA-normalization in that sense would keep the LOS at a constant distance, and only changes in the other components of the PDP could lead to non-stationarities. 

\section{Results}

Based on the evaluation procedure described in the previous section, we evaluate all 1,174 measurement positions that were recorded in our campaign, covering a total of 874 MIMO snapshots for the convoy driving scenario, and 150\,MIMO snapshots for each of overtaking and driving in opposite directions. The results obtained are summarized below. 
      
\begin{figure*}[!t]
        \centering
        \includegraphics[width=6in]{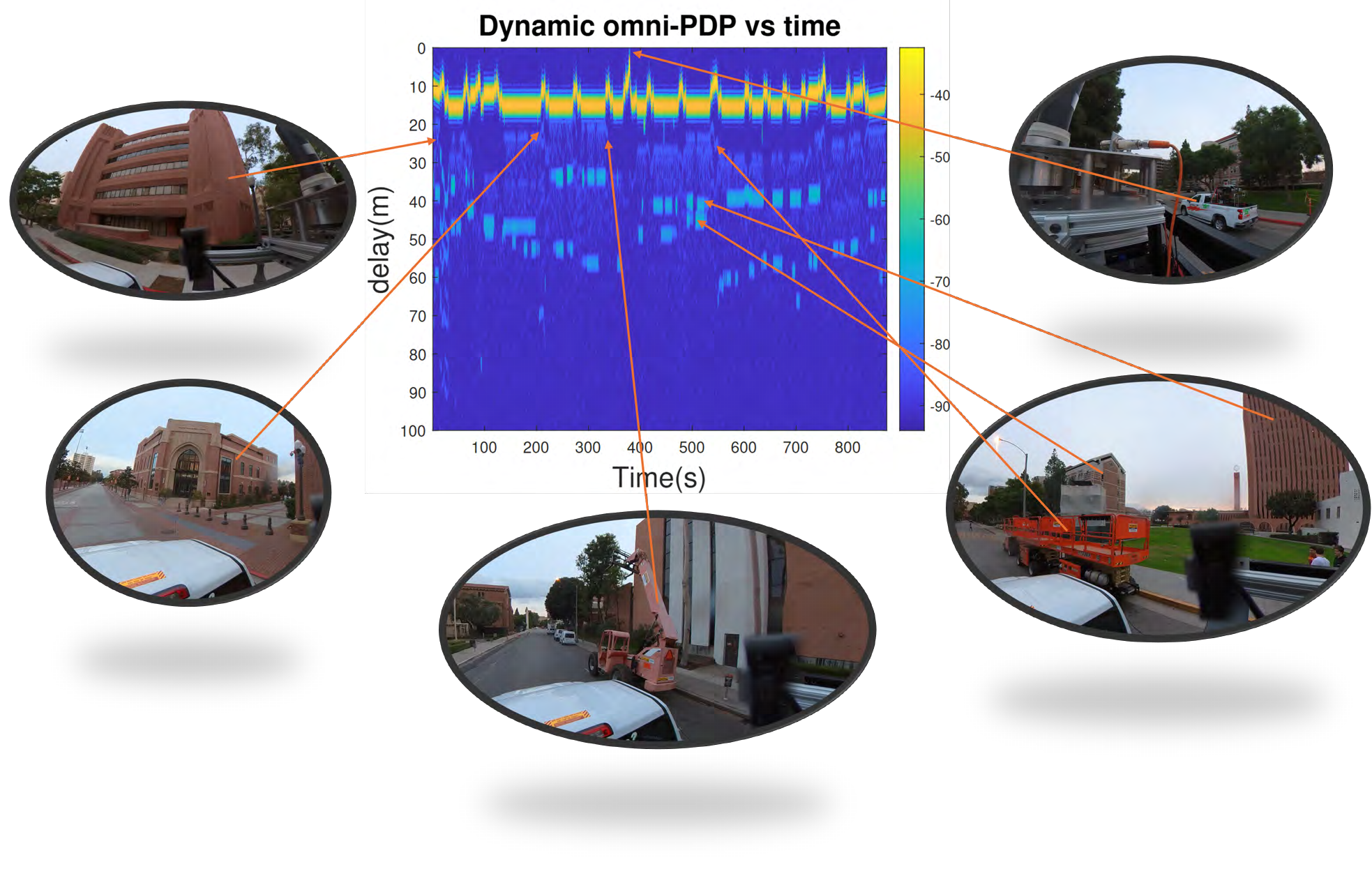}
        \caption{Convoy scenario: time-variant omnidirectional PDP. Main reflectors were identified as buildings (Gerontology, Physical Education, Leavy Library), cranes and trucks. Upper left sub-figure shows the time when the TX-RX distance was minimal because of the U-turn. Other components are traceable aswell but were omitted for visibility and space reasons. }
        \label{Examples_convoy}
        \end{figure*}

\subsection{Power delay profiles}

\begin{figure}[!t]
        \centering
        \includegraphics[width=\columnwidth]{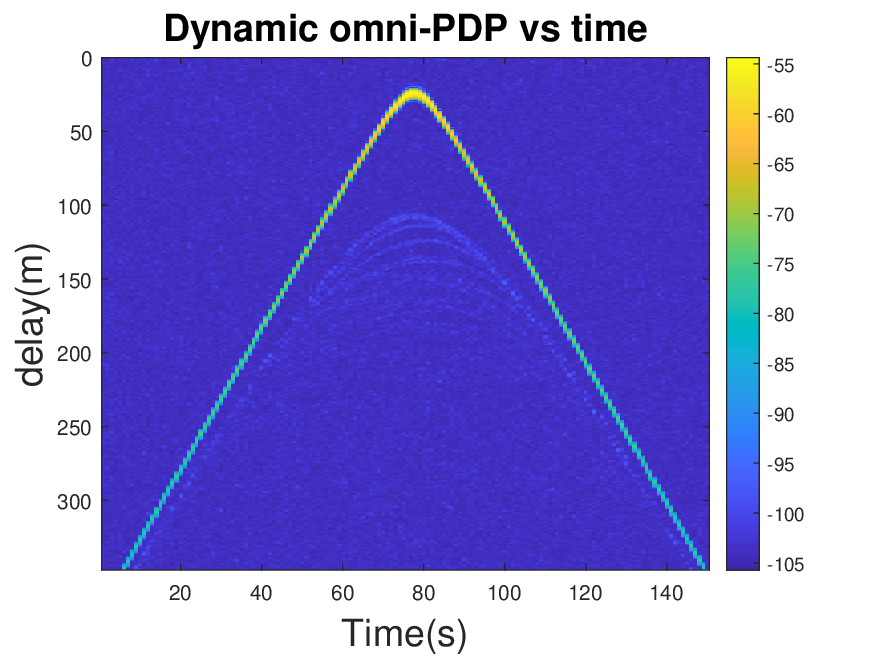}
        \caption{Opposide side scenario: time-variant omnidirectional PDP}
\label{PDP_omni_time_opposite_sides}
        \end{figure}

\begin{figure*}[!t]
        \centering
        \includegraphics[width=6in]{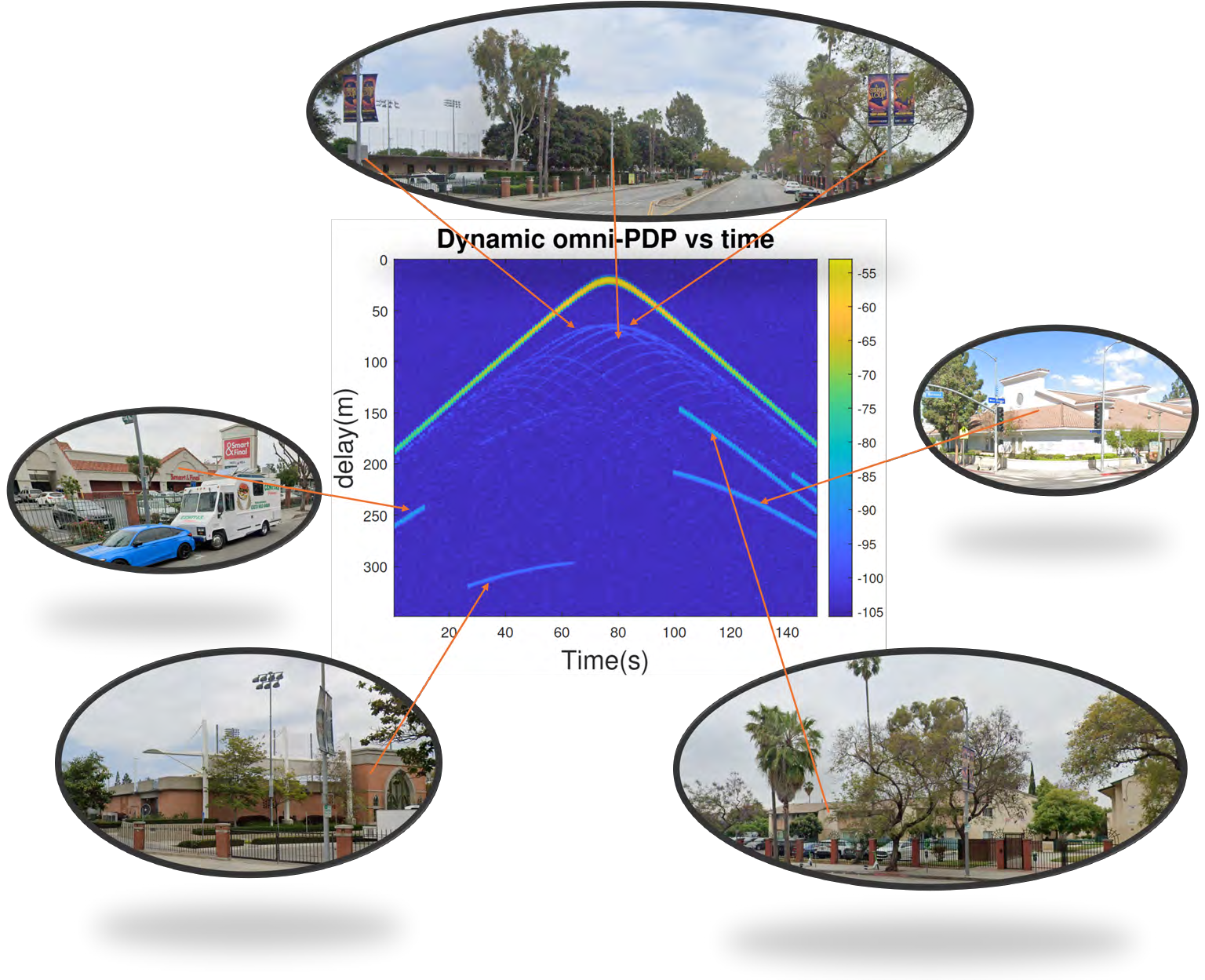}
        \caption{Overtaking scenario: time-variant omnidirectional PDP. Main reflectors identified as lamp-posts, the Smart\&Final market, the united states postal service (USPS) building, the mail-stop processing building at USC, and some residential complexes.}
        \label{Examples_overtaking}
        \end{figure*}

We first show the omnidirectional PDP. To facilitate interpretation, all delays are multiplied by the speed of light so that they correspond to distance. The dynamic evolution of these PDPs vs time are shown in Figs. \ref{Examples_convoy}, \ref{PDP_omni_time_opposite_sides} and \ref{Examples_overtaking}. 

Looking at the convoy driving scenario first, we clearly observe the attempt at keeping a constant separation distance between the two vehicles. We also observe that in comparison to Fig. \ref{Convoy_Map}, the positions at which the vehicles stopped (marked by the arrow-ending points on the map) correspond to the time instants for which the TX-RX distance significantly decreased. The minimum TX-RX separation distance was seen at t = 370\,s, corresponding to the instant at which the cars undergo the U-turn as can be seen in Fig. \ref{Convoy_Map}. In general, line-of-sight (LOS) was maintained throughout the measurement, and separation distance ranged between 4\,m and 18\,m. In addition to the LOS component, we observe significant MPCs with delays up to 65\,m that were traced back to be surrounding buildings and vehicles high enough to be captured by our antenna beam (height of two meters), as confirmed by the video recordings. An example mapping between surrounding objects and the MPCs seen in the PDPs is shown in Fig. \ref{Examples_convoy}. Worth to note here is that other components seen in the PDPs can be traced and explained from geometry as well, however they were omitted for space reasons. 

Next we look at the overtaking and driving on opposite sides scenarios as described in Fig. \ref{other_map}. We observe major similarities in the dynamic omnidirectional PDPs between the two scenarios as can be seen in Figs. \ref{PDP_omni_time_opposite_sides} and \ref{Examples_overtaking}, which is expected given the similarity in the environment for the two scenarios. In general, LOS has been maintained throughout the measurement duration for both scenarios. We also see the LOS starting at high delay for either scenario, then decreasing monotonically at a rate matching the vehicles' speeds until it hits a minimum at time t = 75\,s, corresponding to the position for which the TX-RX separation is minimized. 
An exemplary mapping between surrounding objects and the MPCs seen in the PDPs for the overtaking scenario is shown in Fig. \ref{Examples_overtaking}. We observe significant contributions (about 20\,dB below LOS power) from the metallic light-posts on the sidewalks placed at regular intervals.  We omit to show an example for the driving on opposite sides case; it is noteworthy that the overtaking scenario, Fig. \ref{Examples_overtaking}, shows a number of strong components that are not visible for the opposite-side scenario, due to the positioning of the vehicles.

\subsection{Angular power profiles}

\begin{figure}[!t]
        \centering
        \includegraphics[width=\columnwidth]{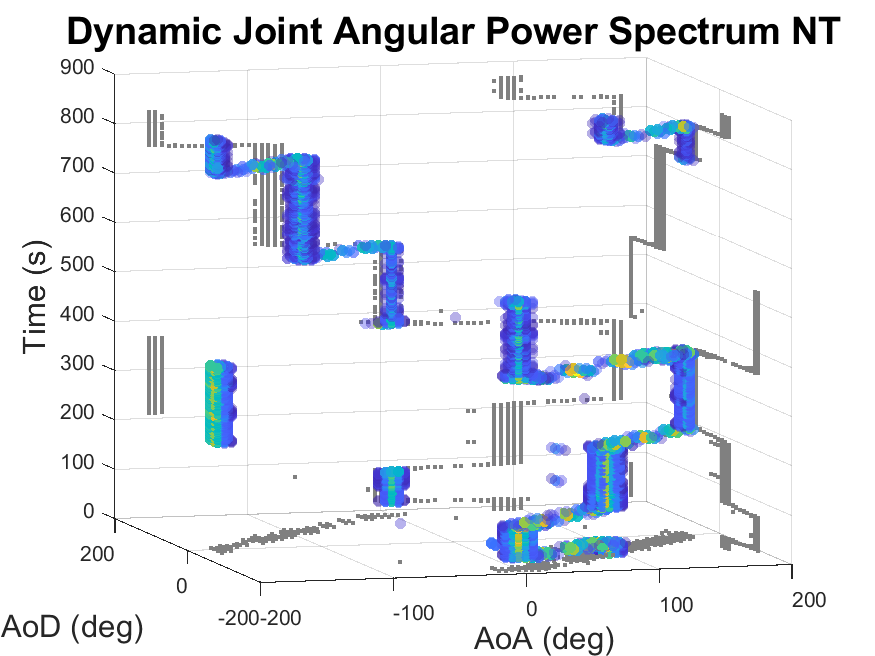}
        \caption{Convoy scenario: time-variant joint APS}
\label{Dynamic_JointAPS_NT_convoy}
        \end{figure}
\begin{figure}[!t]
        \centering
        \includegraphics[width=\columnwidth]{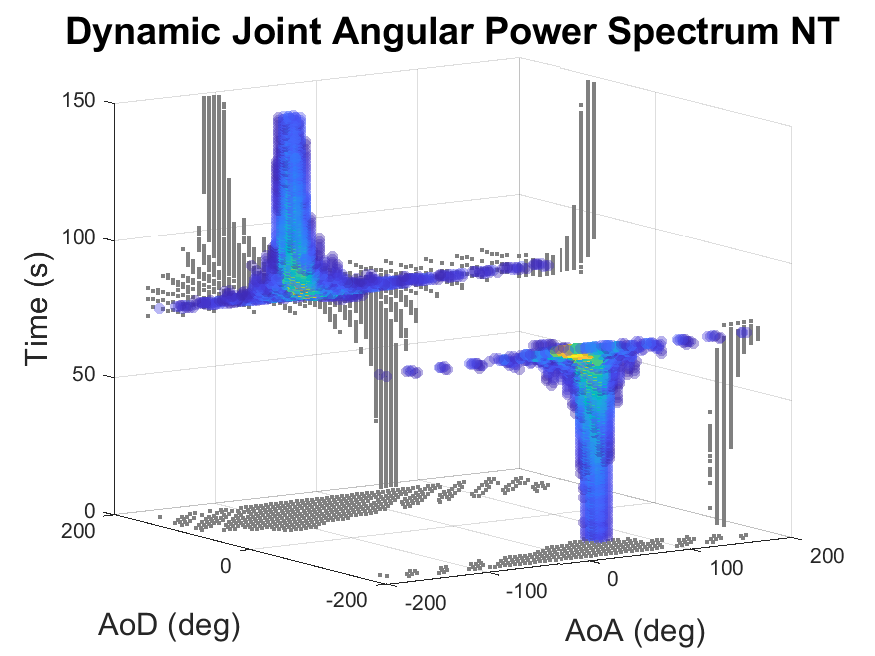}
        \caption{Opposite sides scenario: time-variant joint APS}
\label{Dynamic_JointAPS_NT_Opposite}
        \end{figure}
\begin{figure}[!t]
        \centering
        \includegraphics[width=\columnwidth]{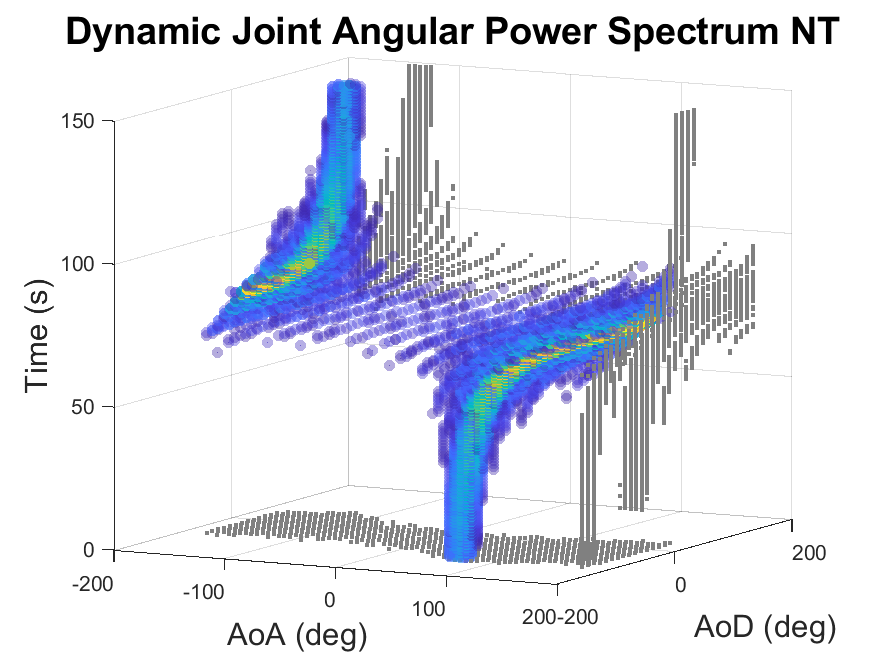}
        \caption{Overtaking scenario: time-variant joint APS}
\label{Dynamic_JointAPS_NT_Overtaking}
        \end{figure}

In terms of APSs, we show the dynamic evolution of the joint TX-RX APSs with time, for all three scenarios in Figs. \ref{Dynamic_JointAPS_NT_convoy}, \ref{Dynamic_JointAPS_NT_Opposite}, and \ref{Dynamic_JointAPS_NT_Overtaking}. We also show the marginals, i.e., the projections of the 3D plot on all three planes for a better visualization of the different contributing MPCs. 

Starting with the convoy scenario in Fig. \ref{Dynamic_JointAPS_NT_convoy}, we clearly observe every turn that was taken by the cars as we proceed in the measurement. We observe limited contribution from MPCs coming from different directions, due to the environment being dominated by the short-distance LOS MPC. Such contributions were tracked back through the video footage and were identified as reflections coming from surrounding trucks and/or buildings. It is worth to note here that an additional threshold of 25\,dB below the maximum was placed on the data, suppressing weak MPCs. This was done to be able to visualize the LOS MPC throughout the duration of the measurement (i.e. other components were removed by the threshold to not crowd the plot or add any ambuiguity in following the track). 

For the driving on opposite sides and the overtaking scenarios, we observe considerable similarity between the dynamic APSs, similar to our observations for the PDPs. 
There is a main cluster of MPCs in the LOS direction, but also major contributions from surrounding lampposts placed regularly on both sides of the street. Additionally, depending on the measurement positions, there are noticeable reflections caused by the United States Postal Service (USPS) building located at the intersection between Vermont and W 36th Street. 

\subsection{Path loss}

\begin{figure}[!t]
        \centering
        \includegraphics[width=\columnwidth]{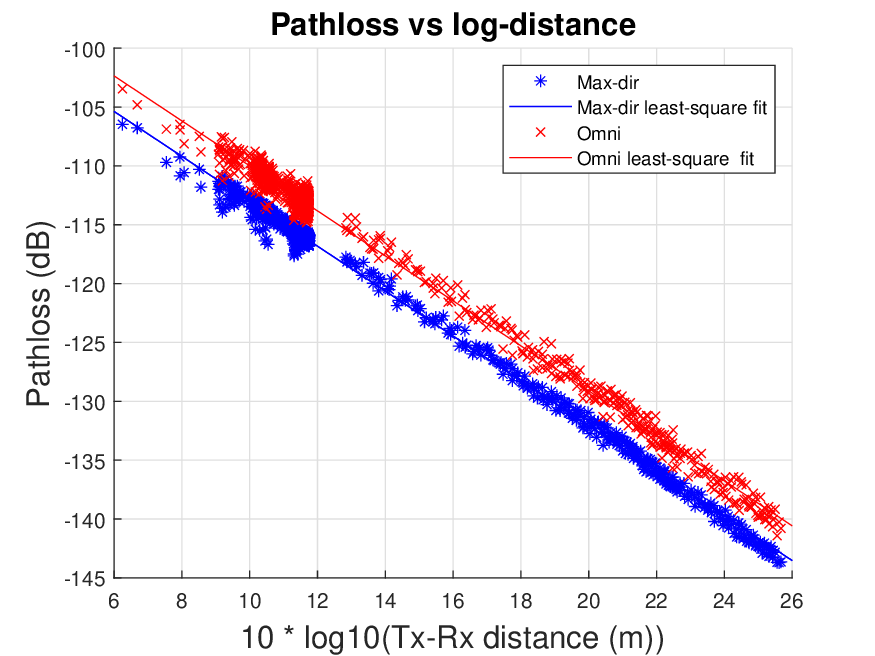}
        \caption{Linear fit for the omnidirectional path loss vs log(d)}
\label{PL_fit}
        \end{figure}

We proceed with the analysis of the statistics (over the ensemble of MIMO snapshots) of the channel parameters, starting with path loss. We plot in Fig. \ref{PL_fit} the path loss values for all positions vs the logarithm of the distance, where the distance was measured as the Euclidean distance of the TX and RX cars whose coordinates are captured by a positioning sensor. The observations for all measurements show a close agreement with free-space propagation model with small variation around the mean. The path loss coefficient evaluated as the slope from the linear fit in Fig. \ref{PL_fit} is at 1.91 for the omnidirectional case and 1.90 for the max-dir case. This is smaller than the FSPL coefficient of 2, which is physically reasonable, as the total received signal consists of several reflected MPCs  in addition to the LOS component that provides power decaying with $d^2$. 


\subsection{RMS delay spread}

\begin{figure}[!t]
        \centering
        \includegraphics[width=\columnwidth]{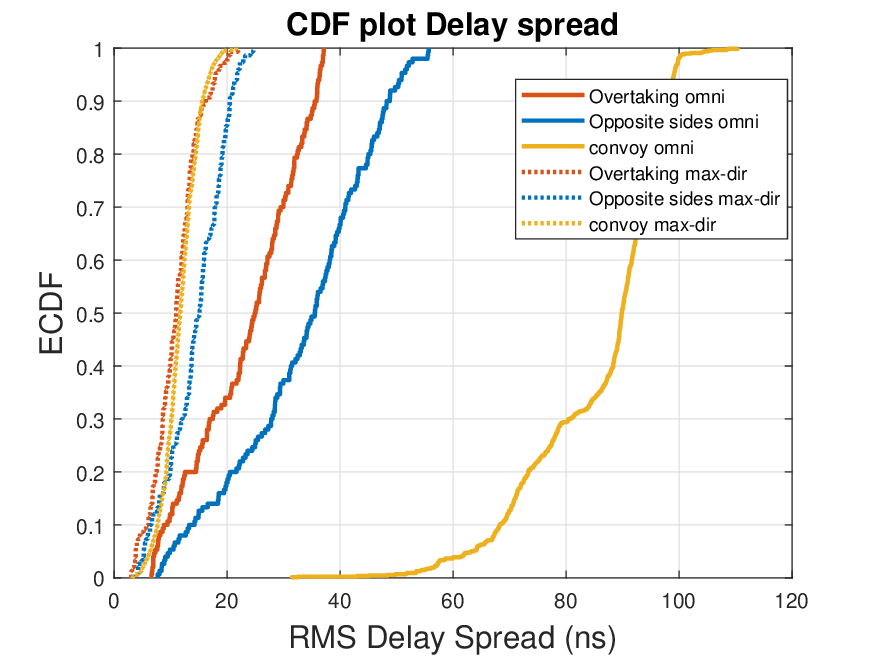}
        \caption{CDF plot of the RMS delay spread in ns}
\label{DS_all}
        \end{figure}

\begin{figure}[!t]
        \centering
        \includegraphics[width=\columnwidth]{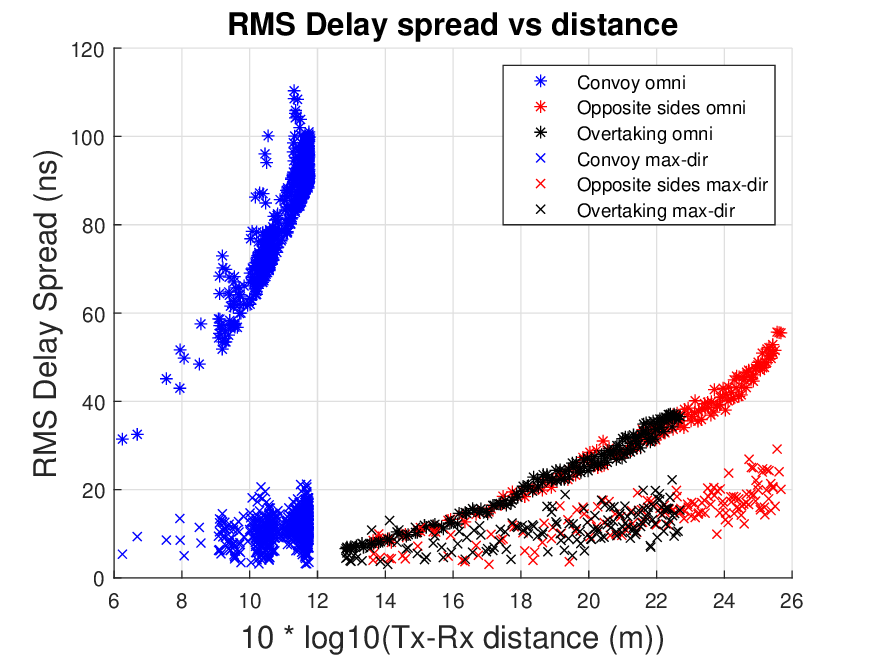}
        \caption{RMS delay spread vs log(distance)}
\label{DS_distance}
        \end{figure}

The RMS delay spread ranges between 5 - 35\,ns for overtaking, 7 - 55\,ns for opposite sides, and 30 - 110\,ns for the driving in convoy scenario. The CDF plots for the three scenarios can be seen in Fig. \ref{DS_all}. We observe significantly higher delay spread for the driving in convoy scenario where we see more contribution from MPCs of varying path-lengths. We conjecture that this is not a matter of the ``convey" driving style, but rather of the different type of street in which the overtaking/opposite driving was happening, with the latter much wider, and fewer buildings and parked cars along the roadside. A plot showing the dependence of the RMS delay spread on the log-distance can be seen in Fig. \ref{DS_distance}. We observe an increase of the delay spread with distance, which is consistent with the well-established results for cellular environments. We also observe that the slope of the increase is considerably larger in the convoy environment than in the opposite/overtaking environment. 

\subsection{Angular spreads}

\begin{figure}[!t]
        \centering
        \includegraphics[width=\columnwidth]{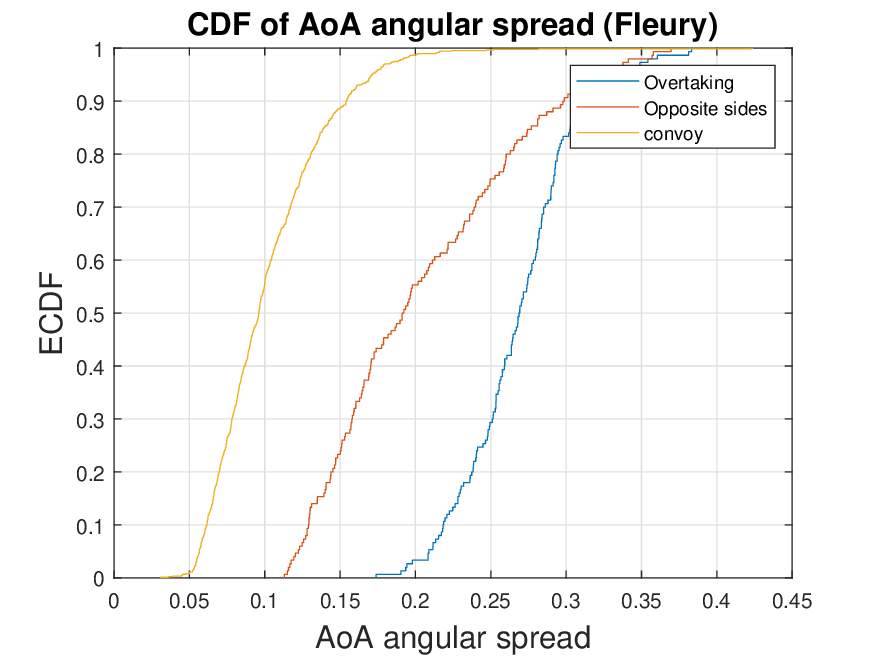}
        \caption{CDF plot of the AoA angular spread in Fleury's definition}
\label{AS_AoA}
        \end{figure}
\begin{figure}[!t]
        \centering
        \includegraphics[width=\columnwidth]{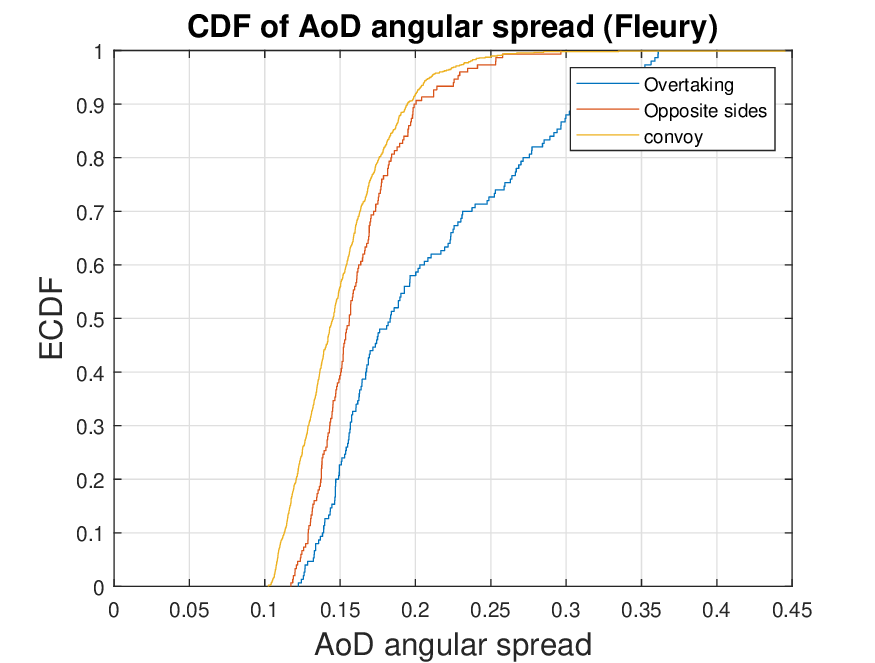}
        \caption{CDF plot of the AoD angular spread in Fleury's definition}
\label{AS_AoD}
        \end{figure}
\begin{figure}[!t]
        \centering
        \includegraphics[width=\columnwidth]{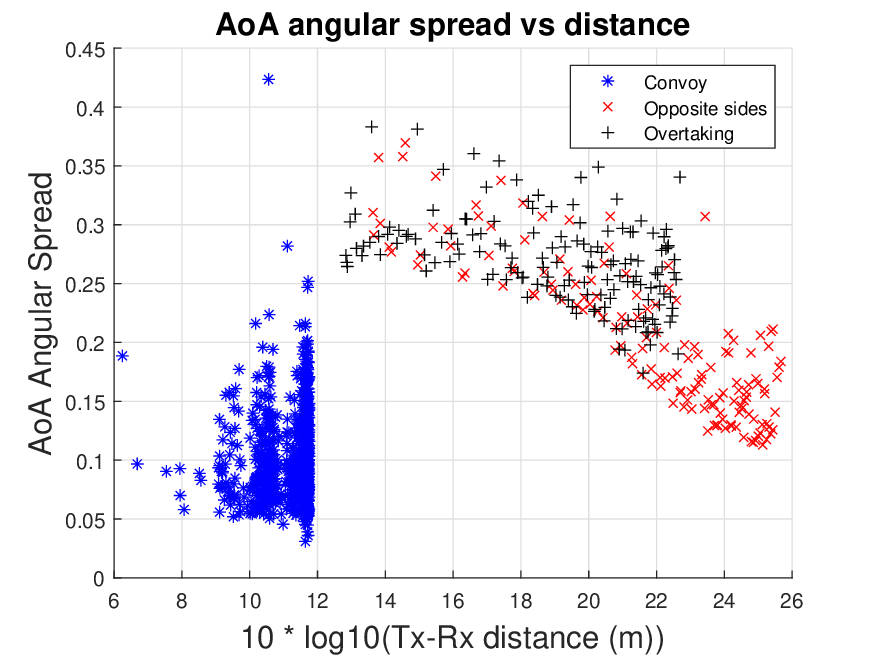}
        \caption{AoA Angular spread vs log(distance)}
\label{AoA_distance}
        \end{figure}
\begin{figure}[!t]
        \centering
        \includegraphics[width=\columnwidth]{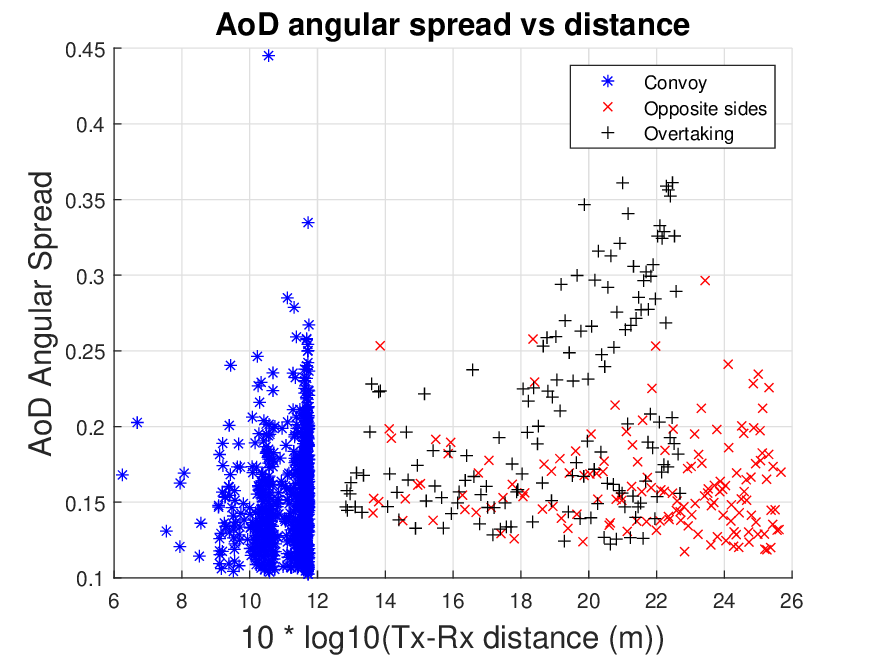}
        \caption{AoD Angular spread vs log(distance)}
\label{AoD_distance}
        \end{figure}

For the statistics of the angular spread, we show the CDF plots for both angles of departures (AoD) and angles of arrivals (AoA) for all three scenarios in Figs. \ref{AS_AoA} and \ref{AS_AoD}. We note relatively low spreads for both the departure and arrival angles which is expected given the measurement scenarios (LOS maintained throughout the measurement). We show the dependence of the angular spreads for both the TX and RX vs the log-distance in Figs. \ref{AoA_distance} and \ref{AoD_distance}. It can be observed that there is relatively little dependence of angular spread on the distance. It is also worth to note that despite this being a peer-to-peer measurement scenario where it is expected for the statistics for the AoA and AoD to be similar, we only observe that similarity in the convoy driving scenario. We speculate that this is due to the fact that only during the convoy scenario the TX and the RX have followed the same path with the same surrounding; for the other two scenarios, the RX was driving on the side of the road that is much more rich in surrounding structures and trees, causing significantly higher angular spreads on the RX. 


\subsection{Power distribution among MPCs}

\begin{figure}[!t]
        \centering
        \includegraphics[width=\columnwidth]{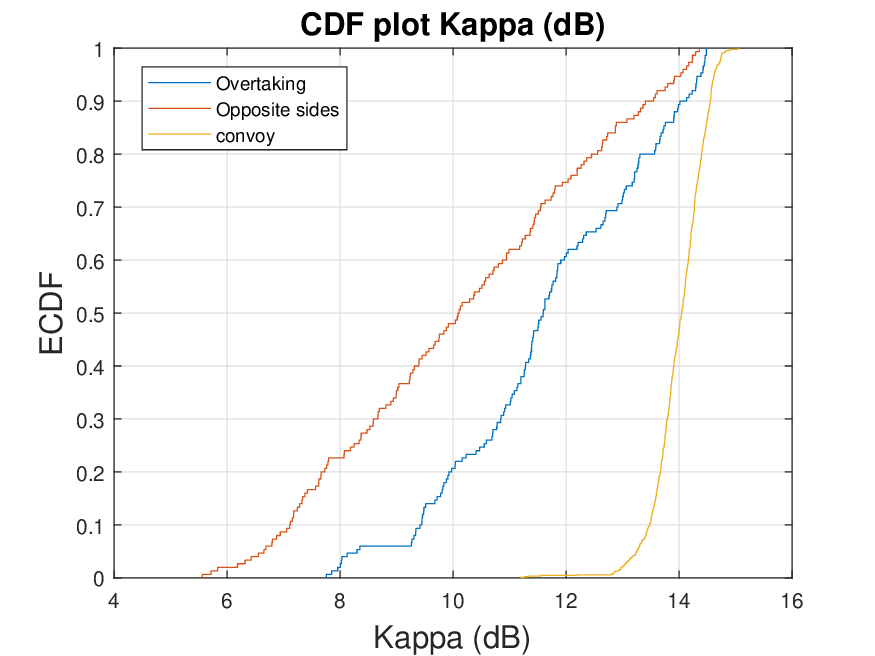}
        \caption{CDF plot of the $\kappa$ parameter in dB}
\label{Kappa}
        \end{figure}
\begin{figure}[!t]
        \centering
        \includegraphics[width=\columnwidth]{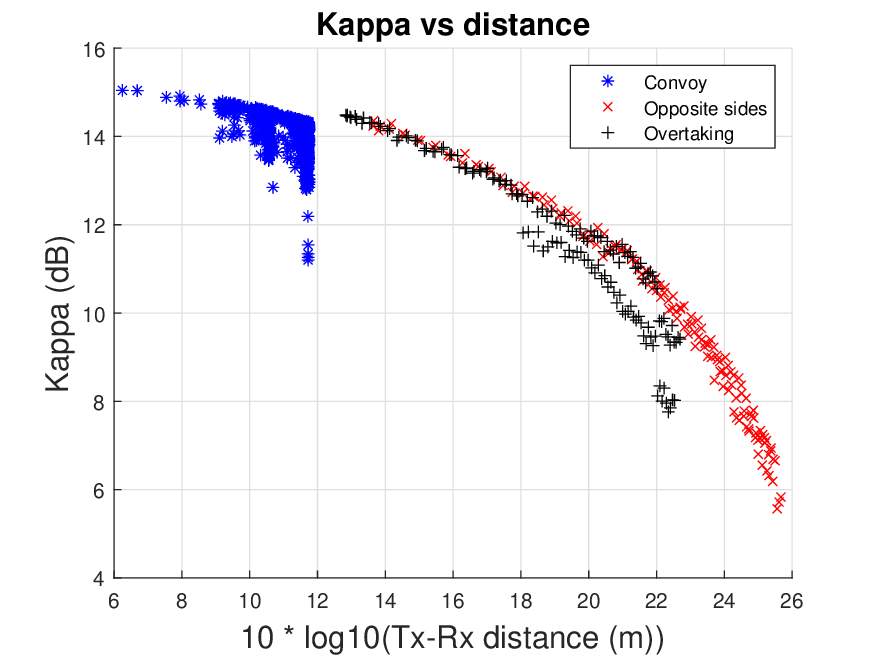}
        \caption{$\kappa$ vs log(distance)}
\label{Kappa_distance}
        \end{figure}

From physical considerations of the measured scenarios, and given what was observed in the previous evaluations about low delay and angular spreads, we expect the values of $\kappa$ to be large (on a dB scale) since the power is mostly concentrated in the LOS component. This is indeed confirmed by the evaluations as can be seen in the $\kappa$ CDF plots in Fig. \ref{Kappa} for all three scenarios. In general, the values of $\kappa$ range between 12 and 15\,dB in the convoy driving scenario where the distance between the TX and RX was mostly maintained constant. Larger ranges are observed for the overtaking and driving on opposite sides scenarios where we see $\kappa$ values ranging between 6 and 14\,dB. A plot showing the dependence of $\kappa$ on the log-distance between the TX and the RX is given in Fig. \ref{Kappa_distance} for all three scenarios. 

\subsection{Stationarity time}

\begin{figure}[!t]
        \centering
        \includegraphics[width=\columnwidth]{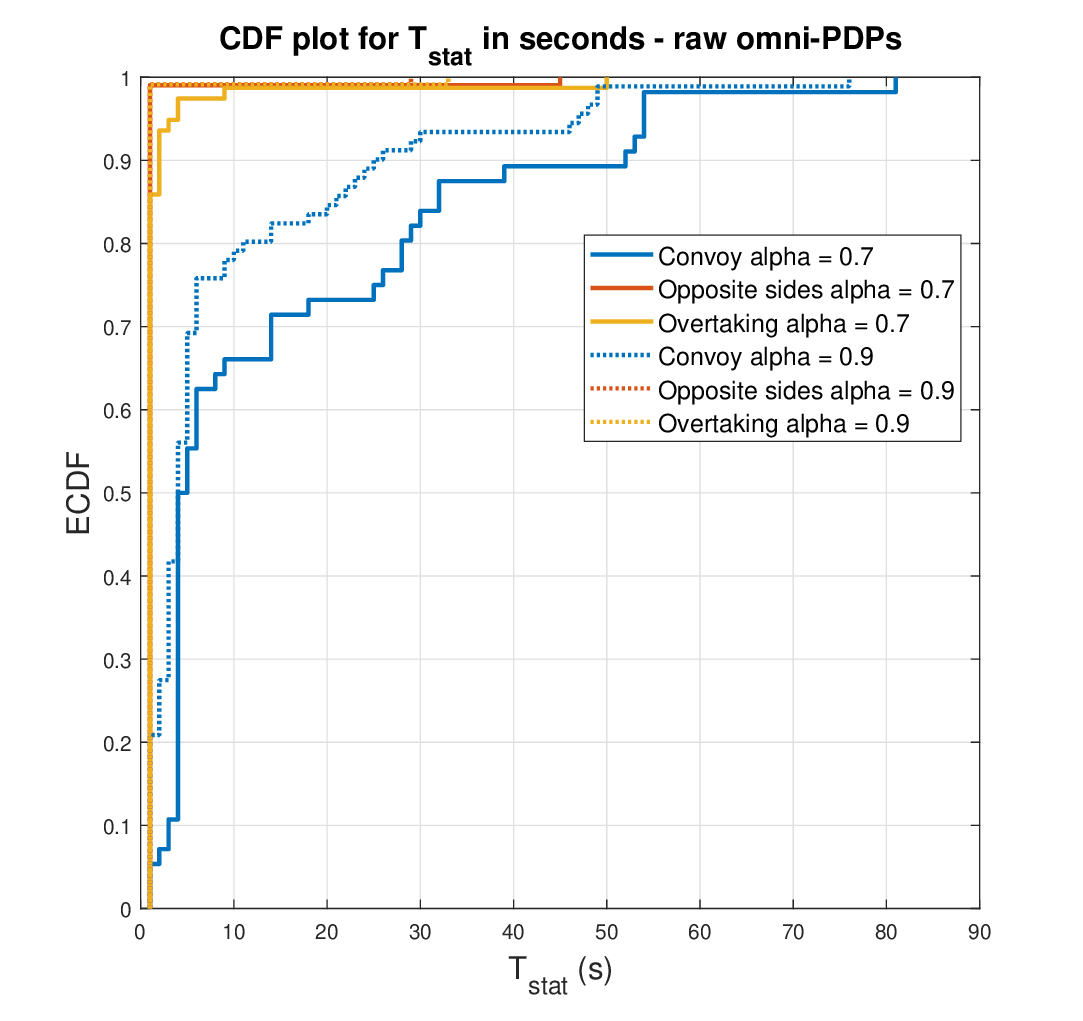}
        \caption{CDF plot of the stationarity time $T_{\rm stat}$ parameter in seconds for the omnidirectional non-normalized PDPs}
\label{Tstat_raw_omni}
        \end{figure}
\begin{figure}[!t]
        \centering
        \includegraphics[width=\columnwidth]{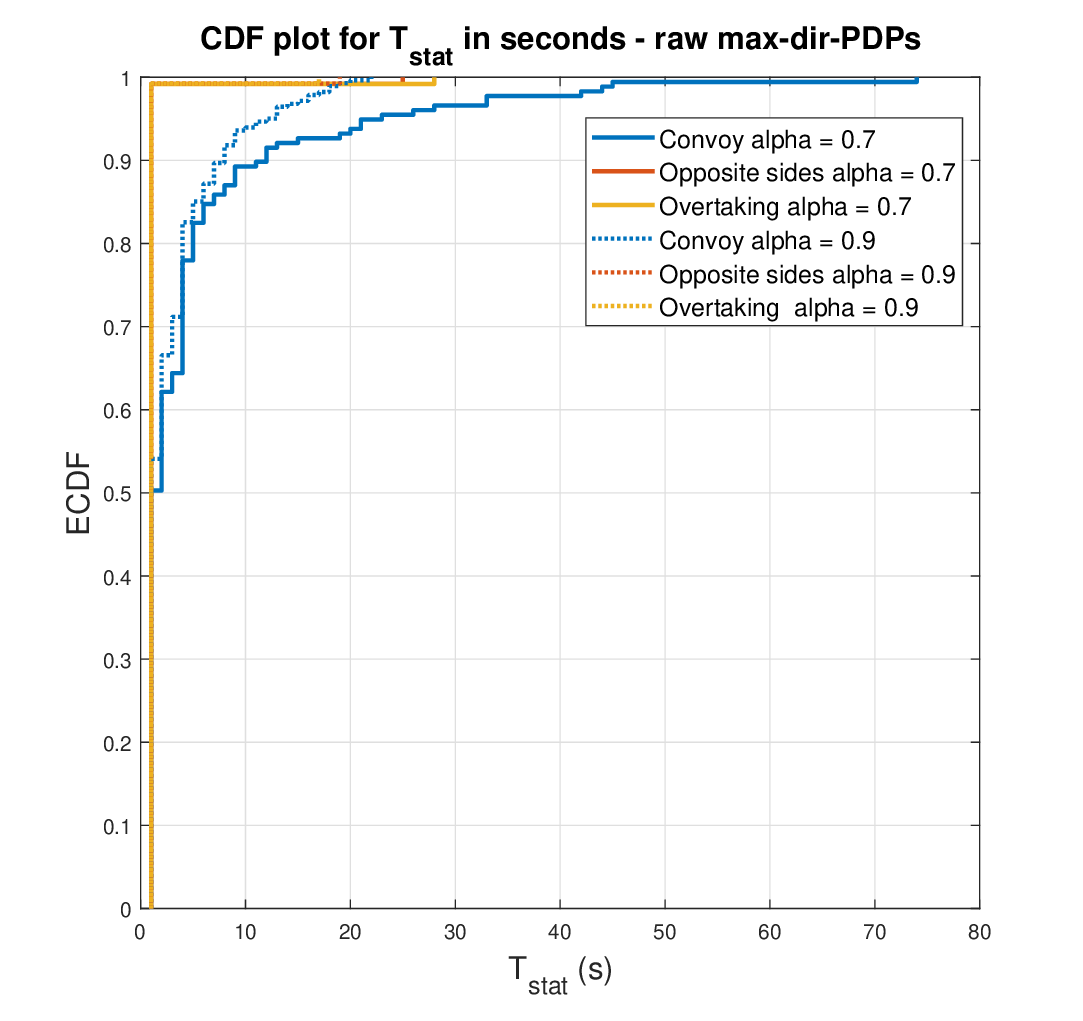}
        \caption{CDF plot of the stationarity time $T_{\rm stat}$ parameter in seconds for the max-dir non-normalized PDPs}
\label{Tstat_raw_maxdir}
        \end{figure}

Finally, we evaluate the stationarity time of the omnidirectional PDP for the three driving scenarios and show the results in the CDF plot for $T_{\rm stat}$ in Fig. \ref{Tstat_raw_omni}, for correlation thresholds $\alpha$ = 0.7 and 0.9. We observe that for the convoy driving scenario, in 50\% of the cases, our evaluated stationarity time is around 5\,s for $\alpha$ = 0.9 threshold, and between 5s and 20s for $\alpha$ = 0.7 threshold. This shows that indeed for our measurement scenarios, our sampling duration falls comfortably within the stationarity region of the PDP of the channel. For the opposite-side and overtaking scenarios, the stationarity times are much shorter, since the delay of the strongest (i.e., LOS) peak changes as the vehicles are moving. As a matter of fact, the stationarity times are almost completely governed by these evolutions, so that the specific values say less about the environment and more about the distance between the vehicles as function of time. We also show the stationarity time of the max-dir PDPs for the three driving scenarios in Fig. \ref{Tstat_raw_maxdir}, again for correlation thresholds $\alpha$ = 0.7 and 0.9. We observe significantly shorter stationarity time for the max-dir PDPs since for this case, and given that our scenarios are all LOS, stationarity is a direct function of the distance between the vehicles and any fluctuation in the distance would have major effect on the correlation between the PDPs.  

\begin{figure}[!t]
        \centering
        \includegraphics[width=\columnwidth]{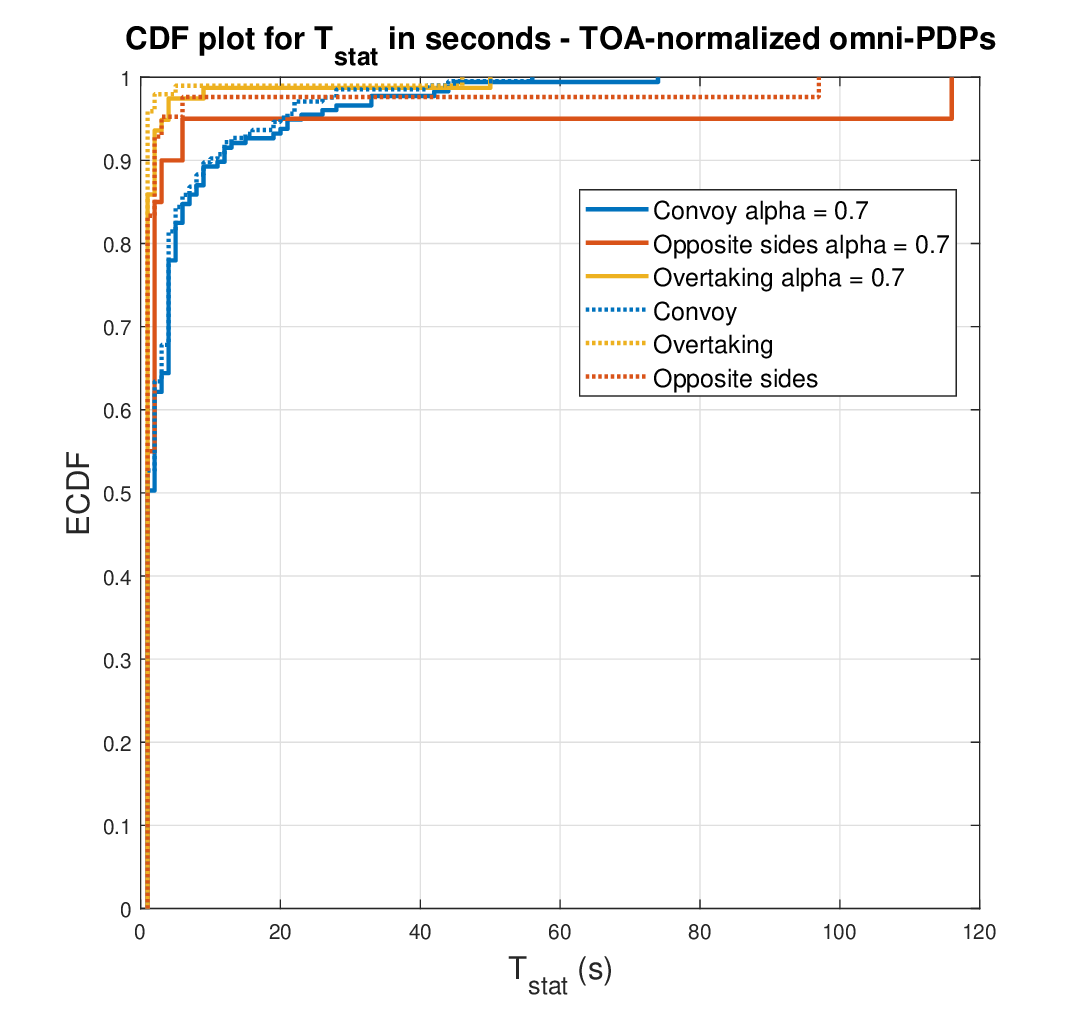}
        \caption{CDF plot of the stationarity time $T_{\rm stat}$ parameter in seconds for the omnidirectional TOA-normalized PDPs}
\label{Tstat_normalized_omni}
        \end{figure}

To be able to visualize the effects of the environment on the stationarity time of the PDPs without such an influence of the LOS component, we additionally provide results for the TOA-normalized PDPs where we normalize the PDPs in delay to the LOS component. We show the results for both omnidirectional and max-dir PDPs for the three driving scenarios and for correlation thresholds $\alpha$ = 0.7 and 0.9. Starting with the omnidirectional PDPs in Fig. \ref{Tstat_normalized_omni}, we see now much larger range for the stationarity time, especially in the opposite sides driving scenario, where we see a stationarity region of length about 120s, roughly 80\% of the measurement scenario duration. We observe shorter $T_{\rm stat}$ values for the convoy scenario, where we see that after normalizing by the LOS delay, surrounding structures start having much larger influence and thus any major changes in the structures surrounding the vehicles causes non-stationarities.

\begin{figure}[!t]
        \centering
        \includegraphics[width=\columnwidth]{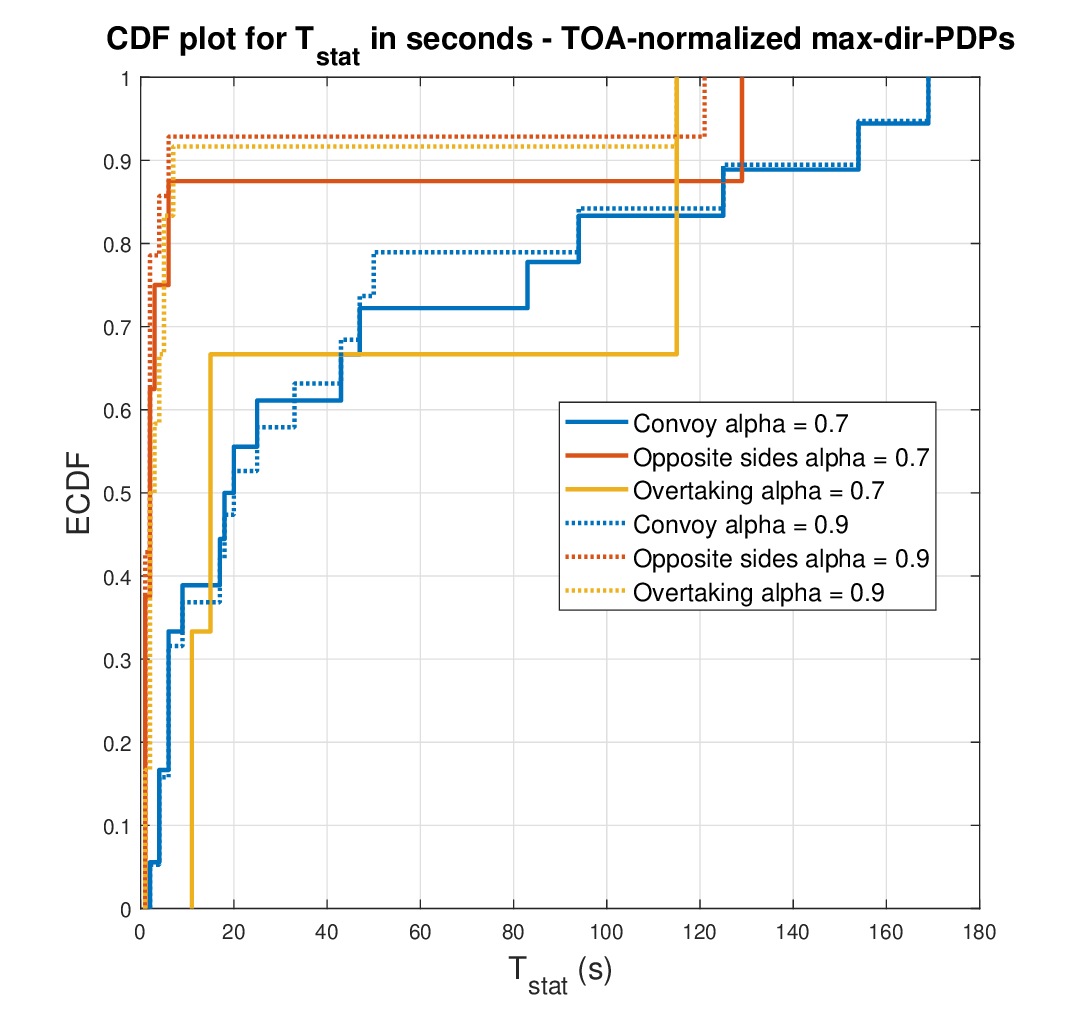}
        \caption{CDF plot of the stationarity time $T_{\rm stat}$ parameter in seconds for the max-dir TOA-normalized PDPs}
\label{Tstat_normalized_maxdir}
        \end{figure}

Finally, we also look at the stationarity time for the TOA-normalized max-dir PDPs and show the results in Fig. \ref{Tstat_normalized_maxdir}, where now that we are looking at only the maximum direction (corresponding to the LOS direction in such scenarios), and after calibrating by the changes in the LOS delay, we observe extremely large values of $T_{\rm stat}$. 

\section{Conclusion}

In this paper, we presented the results of an extensive double-directional measurement campaign for mm-wave V2V channels, using the ReRoMA sounding principle. We have provided an overview of the measurement methodology and environments, as well as the signal processing done to extract parameters of interest in V2V wireless communication system design. We have measured the dynamic channel at 60\,GHz in convoy, overtaking, and driving in opposite directions scenarios. We have observed path loss with a path loss coefficient $n_{\rm omni}$ = 1.91 and $n_{\rm max-dir}$ = 1.90, RMS delay spread ranges in 5-120\,ns and 3-25\,ns for omni and max-dir respectively, angular spread values between 0.05 and 0.4 under Fleury's unitless definition (corresponding to roughly 3-23\,degree spreads). These values are compared to existing mm-wave literature in Table \ref{table:comparison}. Additionally, we have evaluated the power distribution among MPCs as a parameter $\kappa$ and found its values ranging between 6 and 15\,dB. And finally, we have evaluated the stationarity time of the PDPs with and without TOA-normalization to the LOS component and shown that our measurement duration falls comfortably within the stationarity time of the channel. To the best of the authors' knowledge, no comparable results are available in the literature. 

\begin{table*}[t]
\centering
\label{table:comparison}
\caption{Comparison of our results to existing mm-wave literature}
\begin{tabular}{|l|l|l|l|l|l|l|l|}
\hline
\textbf{Author}             & \textbf{Sounder type}   & \textbf{Carrier Freq.} & \textbf{Path loss coeff} & \textbf{Scenario}          & \textbf{Delay Spread}             & \textbf{AoA spread}   & \textbf{AoD spread}   \\ \hline
\textbf{Our work}    & ReRoMA             & 60 GHz                 & 1.90                      & Urban                      & \makecell{5-120ns omni \\  3-25ns max-dir}                            &       3-23deg               &     3-20deg                \\ \hline
\textbf{Hoellinger et al.} \cite{hoellinger2022v2v} & Omnidirectional       & 26 GHz                 & 1.92                     & Urban                      & 51 ns                             & N/A                   & N/A                   \\ \cline{5-6}
                            &                       &                        &                          & Suburban                   & 80 ns                             & N/A                   & N/A                   \\ \hline

\textbf{Zöchmann et al.} \cite{zochmann2018statistical}   & Omni/ Fixed Horn             & 60 GHz                 & N/A                      & Urban                      & 1-4 ns                            &        N/A               &      N/A                 \\ \hline

\textbf{Uyrus et al.}  \cite{uyrus2019visible}     & Fixed Horn                  & 26.5 GHz               & 1.49                     & Parking garage             & N/A                               & N/A                   &       N/A                \\ \cline{3-4}
                            &                       & 38.5 GHz               & 1.58                     & Parking garage             & N/A                               & N/A                   &                       \\ \hline

\textbf{Park et al.}  \cite{park2018vehicle}      & Omni/Fixed Horn             & 28 GHz                 & N/A                      & Parking Lot                & 0.17-3.21 ns                      & 7.03-9.09 deg         &                       \\ \hline

\textbf{Chopra (AT\&T) et al.} \cite{chopra2022real} & Phased Arrays      & 28 GHz                 & 1.77                     & Urban                      & N/A                               & N/A                   & N/A                   \\ \hline
                            &                       & 30 GHz                 & N/A                      & Urban                      & 19 ns                             & 27.3 deg              & 26.9 deg              \\ \cline{3-8}
\textbf{Boban et al.}  \cite{boban2019multi}     & Rotating Horn         & 60 GHz                 & N/A                      & Urban                      & 16.3 ns                           & 19.7 deg              & 19 deg                \\ \cline{3-8}
                            &                       & 73 GHz                 & N/A                      & Urban                      & 10.2 ns                           & 25.6 deg              & 25.1 deg              \\ \cline{3-8}
                            &                       & 73 GHz                 & N/A                      & Highway                    & 6.1 ns                            & 23.4 deg              & 23 deg                \\ \hline

\textbf{Bas et al.}   \cite{bas2017real, bas2018dynamic}      & Phased array          & 27.85 GHz              & 2.051                    & Urban (Campus)             & 50-100 ns                         & 18-26 deg             & 14-23 deg             \\ \hline
\textbf{Yamamoto et al.} \cite{yamamoto2008path}   & Fixed Horn                  & 59.1 GHz               & 0.4 - 1.8                & Highway                    & N/A                               & N/A                   & N/A                   \\ \hline
\textbf{Takahashi et al.} \cite{takahashi2003distance}  & Fixed Horn                  & 60 GHz                 & 1.7 - 2.2                & Highway                    & N/A                               & N/A                   & N/A                   \\ \hline

\textbf{Wang et al.}   \cite{wang2018fading}     & Fixed Horn                  & 73 GHz                 & 2.7                      & Urban                      & N/A                               & N/A                   & N/A                   \\ \hline
\textbf{Sanchez et al.}  \cite{sanchez2017millimeter}  & Fixed Horn                  & 38 GHz                 & N/A                      & Urban                      & 6.52 ns                           & N/A                   & N/A                   \\ \cline{3-8}
                            &                       & 60 GHz                 & N/A                      & Urban                      & 5.92 ns                           & N/A                   & N/A                   \\ \hline
\end{tabular}
\end{table*}

These results can serve for the establishment of double-directional channel models for 5G/6G V2V communications systems at mm-wave frequencies, which in turn enable wireless system designers to exploit delay and spatial diversity and/or multi-stream communications. 

\section{Acknowledgments}
We would like to thank the METRANS project, and in particular Genevieve Giuliano, Cort Brinkerhoff and Katrina Soriano for their kind help in handling many administrative issues over the duration of the project. We would like to also thank the WiDeS@USC group members, especially Guillermo Castro and Jorge Gomez-Ponce, for their help in the hardware design and measurement campaign.

\bibliographystyle{IEEEtran}
\bibliography{mmwave}

\end{document}